\documentclass[12pt]{article}
\usepackage[affil-it]{authblk}
\usepackage[english]{babel} 
\usepackage[T1]{fontenc}
\mathchardef\hyphenmathcode=\mathcode`\-
\usepackage[toc,page,title,titletoc]{appendix}
\usepackage{listings}
\usepackage{xr-hyper}
\usepackage{hyperref,bookmark}
\hypersetup{
  colorlinks=true,
  linkcolor=blue,
  citecolor=blue,
  filecolor=blue,
  urlcolor=blue}
\usepackage{amsbsy}
\usepackage{amsfonts}
\usepackage{amsmath}
\usepackage{amssymb}
\usepackage{amsthm}
\usepackage[toc,page,title,titletoc]{appendix}
\usepackage[fixlanguage]{babelbib}
\usepackage[pdftex]{color}
\usepackage{dsfont}
\usepackage{enumerate}
\usepackage{esvect}
\usepackage[labelfont=bf]{caption}
\usepackage{float}
\usepackage[Glenn]{fncychap}
\usepackage{geometry, calc, color, setspace}%
\usepackage{graphicx,psfrag,epsf}
\usepackage{indentfirst}
\usepackage{lipsum}
\usepackage{longtable}
\usepackage{mathtools}
\usepackage{multirow}
\usepackage[sectionbib]{natbib}
\newtheorem{theorem}{Theorem}[section]

\newtheorem{remark}[theorem]{Remark}
\usepackage{setspace}
\setcitestyle{authoryear, open={(},close={)}}
\usepackage[figuresright]{rotating}
\usepackage{tikz}
\usepackage{varwidth}
\usepackage{authblk}

\usepackage{url} 

\protect
 
\providecommand{\arctanh}{} \renewcommand{\arctanh}{\hspace{2pt}\mathrm{arctanh}}

 \def\@testdef #1#2#3{%
   \def\reserved@a{#3}\expandafter \ifx \csname #1@#2\endcsname
  \reserved@a  \else
 \typeout{^^Jlabel #2 changed:^^J%
 \meaning\reserved@a^^J%
 \expandafter\meaning\csname #1@#2\endcsname^^J}%
 \@tempswatrue \fi}

\begin{document}

\title{\bf A Generalized Heckman Model With Varying Sample Selection Bias and Dispersion Parameters}


\author[1,3]{Fernando de S. Bastos\thanks{E-mail: \texttt{fernando.bastos@ufv.br}}}
\affil[1]{\it Instituto de Ci\^encias Exatas e Tecnol\'ogicas, Universidade Federal de Vi\c cosa, Florestal, Brazil}

\author[2,3]{Wagner Barreto-Souza\thanks{E-mail: \texttt{wagner.barretosouza@kaust.edu.sa}}}
\affil[2]{\it Statistics Program, King Abdullah University of Science and Technology, Thuwal, Saudi Arabia} 
\affil[3]{\it Departamento de Estat\'istica, Universidade Federal de Minas Gerais, Belo Horizonte, Brazil}

\author[2]{Marc G. Genton\thanks{E-mail: \texttt{marc.genton@kaust.edu.sa}}}
 
\date{}
 
\maketitle

\begin{abstract}
Many proposals have emerged as alternatives to the Heckman selection model, mainly to address the non-robustness of its normal assumption. The 2001 Medical Expenditure Panel Survey data is often used to illustrate this non-robustness of the Heckman model. In this paper, we propose a generalization of the Heckman sample selection model by allowing the sample selection bias and dispersion parameters to depend on covariates. We show that the non-robustness of the Heckman model may be due to the assumption of the constant sample selection bias parameter rather than the normality assumption. Our proposed methodology allows us to understand which covariates are important to explain the sample selection bias phenomenon rather than to only form conclusions about its presence. We explore the inferential aspects of the maximum likelihood estimators (MLEs) for our proposed generalized Heckman model. More specifically, we show that this model satisfies some regularity conditions such that it ensures consistency and asymptotic normality of the MLEs. Proper score residuals for sample selection models are provided, and model adequacy is addressed. Simulated results are presented to check the finite-sample behavior of the estimators and to verify the consequences of not considering varying sample selection bias and dispersion parameters. We show that the normal assumption for analyzing medical expenditure data is suitable and that the conclusions drawn using our approach are coherent with findings from prior literature. Moreover, we identify which covariates are relevant to explain the presence of sample selection bias in this important dataset. 
\end{abstract}

\noindent%
\vfill

\doublespacing

\section{Introduction}
\label{sec:intro}

\noindent \citet{heckman1974,heckman1976} introduced a model for dealing with the sample selection bias problem with the aid of a bivariate normal distribution to relate the outcome of interest and a selection rule. A semiparametric alternative to this model, known as Heckman’s two-step method, was proposed by \cite{heckman1979}, so that the non-robustness of the normal distribution in the presence of outliers could be handled.

The most discussed problem regarding the Heckman model is its sensitivity to the normal assumption of errors. Misspecification of the error distribution leads to inconsistent maximum likelihood estimators, yielding biased estimates \citep{TsayPin}. On the other hand, when the error terms are correctly specified, the estimation by maximum likelihood or by procedures based on likelihood produces consistent and efficient estimators \citep{leung1996choice,enders2010applied}.

However, even when the shape of the error density is correctly specified, the heteroskedasticity of the error terms can cause inconsistencies in the parameter estimates, as shown by \cite{hurd1979estimation} and \cite{arabmazar1981further}. In response to this concern, \cite{donald1995two} discussed how heteroskedasticity in sample selection models is relatively neglected and provided two reasons to motivate the importance of taking this into account in practice. The first reason is that typically, the data used to fit sample selection models comprises large databases, in which heterogeneity is commonly found. The second reason is that the estimates of the parameters obtained by fitting the usual selection models may in some cases be more severely affected by heteroscedasticity than by incorrect distribution of the error terms \citep{powell1986symmetrically}.

Nevertheless, even though there is a large body of recent research and studies on sample selection models, few studies have been carried out to correct or minimize the impact of heteroscedasticity. This is one of the aims of our paper.
\cite{Siddhartha} proposed a semiparametric model for data with sample selection bias. They considered nonparametric functions in their model, which allowed great flexibility in the way covariates affect response variables. They still presented a Bayesian method for the analysis of such models. Subsequently, \cite{Wiesenfarth} introduced another general estimation method based on Markov Chain Monte Carlo simulation techniques and used a simultaneous equation system that incorporates Bayesian versions of  penalized smoothing splines. 

Recent works on sample selection models have aimed to address robust alternatives to the Heckman model. In this direction, \cite{marchenkoGenton} proposed a Student-$t$ sample selection model for dealing with the robustness to the normal assumption in the Heckman model. \cite{zhelonkin2013robustness} proposed a modified robust semiparametric alternative based on Heckman's two-step estimation method. They proved the asymptotic normality of the proposed estimators and provided the asymptotic covariance matrix. To deal with departures from normality due to skewness, \cite{ogundimu2016sample} introduced the skew-normal sample selection model to mitigate the remaining effect of skewness after applying a logarithmic transformation to the outcome variable.

Another direction that has been explored in the last few years is the modeling of discrete data with sample selection. For instance, \cite{MARRA2016110} introduced sample selection models for count data, potentially allowing for the use of any discrete distribution, non-Gaussian dependencies between the selection and outcome equations, and flexible covariate effects. The modeling of zero-inflated count data with sample selection bias is discussed by \cite{Wysmar2018}. \cite{Beili} considered the semiparametric identification and estimation of a heteroscedastic binary choice model with endogenous dummy regressors, and no parametric restriction on the distribution of the error term was assumed. This yields general multiplicative heteroscedasticity in both selection and outcome equations and multiple discrete endogenous regressors. A class of sample selection models for discrete and other non-Gaussian outcomes was recently proposed by \cite{azzalinietal}.

\cite{Giampiero} introduced a generalized additive model for location, scale and shape, which accounts for non-random sample selection, and \cite{Taeyoung} proposed a Bayesian methodology to correct the bias of estimation of the sample selection models based on a semiparametric Bernstein polynomial regression model that incorporates the sample selection scheme into a stochastic monotone trend constraint, variable selection, and robustness against departures from the normality assumption.

In the aforementioned papers, the solution to deal with departures from normality for continuous outcomes is to assume robust alternatives like Student-$t$ or skew-normal distributions. Another common approach is to consider non-parametric structures for the density of the error terms. Our proposal goes in a different direction, one which has not yet been explored in the literature. 

In this paper, we propose a generalization of the Heckman sample selection model by allowing the sample selection bias and dispersion parameters to depend on covariates. We show that the non-robustness of the Heckman model may be due to the assumption of the constant sample selection bias parameter rather than the normality assumption. Our proposed methodology allows us to understand what covariates are important to explain the sample selection bias phenomenon rather than to solely make conclusions about its presence. It is worthy to mention that our methodology can be straightforwardly adapted for existing sample selection like those proposed by \cite{marchenkoGenton}, \cite{ogundimu2016sample}, and \cite{Wysmar2018}. We now highlight other contributions of our paper:

\begin{itemize}
    \item We explore the inferential aspects of the maximum likelihood estimators (MLEs) for our proposed generalized Heckman model. More specifically, we show that this model satisfies regularity conditions so that it ensures consistency and asymptotic normality of the MLEs. In particular, we show that the Heckman model satisfies the regularity conditions, which is a new finding.
    
    \item Proper residual for sample selection models is proposed as a byproduct of our asymptotic analysis. This is another relevant contribution of our paper, since this point has not yet been thoroughly addressed.
    
    \item We develop an \texttt{R} package for fitting our proposed generalized Heckman model. It also includes the Student-$t$ and skew-normal sample selection models, which have not been implemented in \texttt{R} \citep{SoftwareR} before. This makes the paper replicable and facilitates the use of our generalized Heckman model by practitioners. 
    
    \item We show that the normal assumption for analyzing medical expenditure data is suitable and that the conclusions drawn using our approach are consistent with findings from prior literature. Moreover, we identify which covariates are relevant for explaining the presence of sample selection bias in this important dataset.
\end{itemize}

This paper is organized as follows. In Section \ref{HeckmanGen} we define the generalized Heckman (GH) sample selection model and discuss estimation of the parameters through the maximum likelihood method. Further, diagnostics tools and residual analysis are discussed. Section \ref{asymptotics} is devoted to showing that the GH model satisfies regularity conditions that ensure consistency and asymptotic normality of the maximum likelihood estimators. In Section \ref{simulations}, we present Monte Carlo simulation results for evaluating the performance of the maximum likelihood estimators of our proposed model and for checking the behavior of other existing methodologies under misspecification. In Section \ref{3:application} we apply our generalized Heckman model to the data on ambulatory expenditures from the 2001 Medical Expenditure Panel Survey and show that our methodology overcomes an existing problem in a simple way. Concluding remarks are addressed in Section \ref{concluding_remarks}. This paper contains a Supplementary Material which can be obtained from the authors upon request.

\section{Generalized Heckman Model}\label{HeckmanGen}

Assume that $\{(Y_{1i}^{*},Y_{2i}^{*})\}_{i=1}^n$ are linearly related to covariates $\pmb{x}_{i}\in \mathbb{R}^{p}$ and $\pmb{w}_{i}\in \mathbb{R}^{q}$ through the following regression structures:
\begin{align}
Y_{1i}^{*}&=\mu_{1i}+\epsilon_{1i}, \label{eq_reg_heckman} \\
Y_{2i}^{*}&=\mu_{2i}+\epsilon_{2i}, \label{eq_sel_heckman} 
\end{align}
where $\mu_{1i}=\pmb{x}_{i}^\top\pmb\beta$, $\mu_{2i}=\pmb w_{i}^\top\pmb\gamma$, $\pmb{\beta}=(\beta_1,\ldots,\beta_p)^\top\in \mathbb{R}^{p}$ and $\pmb{\gamma}=(\gamma_1,\ldots,\gamma_q)^\top\in \mathbb{R}^{q}$ are vectors of unknown parameters with associated covariate vectors $\pmb x_i$ and $\pmb w_i$, for $i=1,\ldots,n$, and $\{(\epsilon_{1i},\epsilon_{2i})\}_{i=1}^n$ is a sequence of independent bivariate normal random vectors. More specifically, we suppose that
\begin{align}\label{2:disterro2}
\begin{pmatrix}
\epsilon_{1i}\\
\epsilon_{2i}
\end{pmatrix} &\sim \mathcal{N}_2
\begin{bmatrix}
\begin{pmatrix}
0\\
0
\end{pmatrix}\!\!,&
\begin{pmatrix}
\sigma_{i}^{2} & \rho_{i}\sigma_{i} \\
\rho_{i}\sigma_{i} & 1
\end{pmatrix}
\end{bmatrix},
\end{align}
with the following regression structures for the sample selection bias and dispersion parameters:
\begin{eqnarray}\label{addreg}
\mbox{arctanh}\,\rho_i=\pmb v_i^\top\pmb\kappa \quad \mbox{and}\quad \log\sigma_i=\pmb e_i^\top\pmb\lambda,
\end{eqnarray}
where $\pmb{\lambda}=(\lambda_{1},\ldots,\lambda_{r})^\top\in \mathbb{R}^{r}$ and $\pmb{\kappa}=(\kappa_{1},\ldots,\kappa_{s})^\top\in \mathbb{R}^{s}$ are parameter vectors, with associated covariate vectors $\pmb v_i$ and $\pmb e_i$, for $ i=1,\ldots,n$. The $\mbox{arctanh}$ (inverse hyperbolic tangent) link function used for the sample selection bias parameter ensures that it belongs to the interval $(-1,1)$. The variable $Y_{1i}^{*}$ is observed only if $Y_{2i}^{*}>0$ while the variable $Y_{2i}^{*}$ is latent. We only know if $Y_{2i}^{*}$ is greater or less than 0. 
The equation (\ref{eq_reg_heckman}) is the primary interest equation and the equation (\ref{eq_sel_heckman}) represents the selection equation. In practice, we observe the variables 
\begin{align}
U_{i}&=I\{Y_{2i}^{*}>0\},\\
Y_{i}&=Y_{1i}^{*}U_{i},\label{mod_heckman_ind}
\end{align}
for $ i=1,\ldots,n$, where $I\{Y_{2i}^{*}>0\}=1$ if $Y_{2i}^{*}>0$ and equals 0 otherwise.  

Our generalized Heckman sample selection model is defined by (\ref{eq_reg_heckman})-(\ref{mod_heckman_ind}). The classic Heckman model is obtained by assuming constant sample selection bias and dispersion parameters in (\ref{addreg}).

The mixed distribution of $Y_{i}$ is composed by the discrete component 
\begin{align}
P(U_i=u_i)=\Phi(\pmb{w}_{i}^\top\boldsymbol{\gamma})^{u_i}\Phi(-\pmb{w}_{i}^\top\boldsymbol{\gamma})^{1-u_i},\quad u_i=0,1,
\end{align}
and by a continuous part given by the conditional density function
\begin{align}
\label{1:dens_heckman}
f(y_{i}|U_i=1)&=\dfrac{1}{\sigma_{i}}\phi\left(\dfrac{y_{i}-\pmb{x}_{i}^\top\boldsymbol{\beta}}{\sigma_{i}}\right)
\dfrac{\Phi\left(\dfrac{\rho_{i}}{\sqrt{1-\rho_{i}^{2}}}\left(\dfrac{y_{i}-\pmb{x}_{i}^\top\boldsymbol{\beta}}{\sigma_{i}}\right)+
\dfrac{\pmb{w}_{i}^\top\boldsymbol{\gamma}}{\sqrt{1-\rho_{i}^{2}}}\right)}{\Phi(\pmb{w}_{i}^\top\boldsymbol{\gamma})},
\end{align}
where $\phi(\cdot)$ and $\Phi(\cdot)$ denote the density and cumulative distribution functions of the standard normal, respectively. More details about this are provided in the Supplementary Material.

Let $\boldsymbol{\theta}=(\pmb{\beta}^{\top},\pmb{\gamma}^{\top},\pmb{\lambda}^{\top},\pmb{\kappa}^{\top})^{\top}$ be the parameter vector. The log-likelihood function is given by 
\begin{align}\label{likelihoodGEN}
\mathcal{L}(\boldsymbol{\theta})&=\sum_{i=1}^{n} \left\{u_{i}\log{f(y_{i}|u_{i}=1)}+u_{i}\log{\Phi(\mu_{2i})}+
(1-u_{i})\log{\Phi(-\mu_{2i})}\right\}\nonumber\\
\displaystyle &=\sum_{i=1}^{n}u_{i}\left\{\log{\Phi(\zeta_{i})}+ \log{\phi(z_{i})}-\log{\sigma_{i}}\right\}+\sum_{i=1}^{n}(1-u_{i})\log{\Phi(-\mu_{2i})},
\end{align}
where $u_{i}=1$ if $y_{i}$ is observed and $u_{i}=0$ otherwise, $ z_{i}\equiv\dfrac{y_{i}-\mu_{1i}}{\sigma_{i}}$, and $\zeta_{i}\equiv\dfrac{\mu_{2i}+\rho_iz_{i}}{\sqrt{1-\rho_i^2}}$, for $i=1,\ldots,n$.

Expressions for the score function ${\bf S}_{\boldsymbol{\theta}}=\dfrac{\partial \mathcal{L}(\boldsymbol{\boldsymbol{\theta}})}{\partial\boldsymbol{\theta}}$ and the respective Hessian matrix are presented in the Supplementary Material. The maximum likelihood estimators (MLEs) are obtained as solution of the non-linear system of equations ${\bf S}_{\boldsymbol{\theta}}={\bf 0}$, which does not have an explicit analytic form. We use the quasi-Newton algorithm of Broyden-Fletcher-Goldfarb-Shanno (BFGS) via the package \texttt{optim} of the software \texttt{R} \citep{SoftwareR} to maximize the log-likelihood function. 

It is important to emphasize that the maximum likelihood method may suffer from possible multicollinearity problems when the selection equation has the same covariates as the regression equation (for example, see \cite{marchenkoGenton}). To reduce the impact of this problem in parameter estimation, the exclusion restriction is suggested in the literature. According to this approach, at least one significant covariate included in the selection equation should not be included in the primary regression. The interested reader can find more details on the exclusion restriction procedure for the Heckman sample selection model in \cite{heckman1976}, \cite{Leung2000} and \cite{Newey2009}.

We now discuss diagnostic techniques, which have been proposed to detect observations that could exercise some influence on the parameter estimates or inference in general. Next, for the generalized Heckman model, we describe the generalized Cook distance (GCD), and in the next section, we propose the score residual.

Cook's distance is a method commonly used in statistical modeling to evaluate changes in the estimated vector of parameters when observations are deleted. It allows us to assess the effect of each observation on the estimated parameters. The methodology proposed by \cite{RDCook} suggests the deletion of each observation and the evaluation of the log-likelihood function without such a case. According to \cite{XIE20074692}, the generalized Cook distance (GCD) is defined by
\begin{align*}
\mathrm{GCD}_{i}(\boldsymbol{\theta})=\left(\widehat{\boldsymbol{\theta}}-\widehat{\boldsymbol{\theta}}_{(i)}\right)^\top\boldsymbol{M} \left(\widehat{\boldsymbol{\theta}}-\widehat{\boldsymbol{\theta}}_{(i)}\right), \quad i=1, \ldots, n,
\end{align*}
where $\boldsymbol{M}$ is a nonnegative definite matrix, which measures the weighted combination of the elements for the difference $\widehat{\boldsymbol{\theta}}-\widehat{\boldsymbol{\theta}}_{(i)}$, and $\widehat{\boldsymbol{\theta}}_{(i)}$ is the MLE of $\boldsymbol\theta$ when removing the $i$th observation. Many choices for $\boldsymbol{M}$ were considered by \cite{cook1982residuals}. We use the inverse variance-covariance matrix $\boldsymbol{M}=-\ddot{\mathcal{L}}(\widehat{\boldsymbol{\theta}})^{-1}.$
To determine whether the $i$th case is potentially influential on inference about $\boldsymbol\theta$, we check if its associated GCD value is greater than $2p/n$. In this case, this point would be a possible influential observation.

We illustrate the usage of the GCD in the analysis of the medical expenditure data in Section~\ref{3:application}. Regarding the residual analysis, in the next section, we propose a proper residual for sample selection models, which is one of the aims of this paper.

\section{Asymptotic Properties and Score Residuals}\label{asymptotics}

Our aim in this section is to show that under some conditions, our proposed generalized Heckman sample selection model satisfies the regularity conditions stated by \cite{cox1979theoretical}. As a consequence, the maximum likelihood estimators discussed in the previous section are consistent and asymptotically normal distributed. As a by-product of our findings here, we propose a score residual that is well-known to be approximately normal distributed. Proofs of the theorems stated in this section can be found in the Appendix.

Let $\boldsymbol\Theta$ be the parameter space and $\ell_i(\boldsymbol\theta)= u_{i}\left\{\log{\Phi(\zeta_{i})}+\log{\phi(z_{i})}-\log{\sigma_{i}}\right\}+(1-u_{i})\log{\Phi(-\mu_{2i})}$ be the contribution of the $i$th observation to the log-likelhood function, where $\zeta_{i}$ retains its definition from previous section, for $i=1,\ldots,n$. 

\begin{theorem}\label{regcond}
The score function associated to the generalized Heckman model has mean zero and satisfies the identity
$E\left({\bf S}_{\boldsymbol{\theta}}{\bf S}_{\boldsymbol{\theta}}^\top\right)=-E\left(\partial {\bf S}^\top_{\boldsymbol{\theta}}/\partial\boldsymbol\theta\right)$.
\end{theorem}

We now propose a new residual for sample selection models inspired from Theorem \ref{regcond}. From (\ref{scorebeta}), we define the ordinary score residual by $s_i=z_i-\dfrac{\rho_i}{\sqrt{1-\rho_i^2}}\dfrac{\phi(\zeta_{i})}{\Phi(\zeta_{i})}$ for the non-censored observations (where $u_i=1$) and the standardized score residual by
\begin{eqnarray*}
S_i=\dfrac{s_i-E(s_i|U_i=1)}{\sqrt{\mbox{Var}(s_i|U_i=1)}}=\dfrac{z_i-\dfrac{\rho_i}{\sqrt{1-\rho_i^2}}\dfrac{\phi(\zeta_{i})}{\Phi(\zeta_{i})}}{\sqrt{1+\mu_{2i}\rho_i^2\dfrac{\phi(\mu_{2i})}{\Phi(\mu_{2i})}+\dfrac{\rho_i^2}{1-\rho_i^2}\Psi_i}},
\end{eqnarray*}
where $\Psi_i=E\left(\dfrac{\phi^2(\zeta_i)}{\Phi^2(\zeta_i)}\big|U_i=1\right)=\dfrac{1}{\sigma_i\Phi(\mu_{2i})}\displaystyle\int_{-\infty}^\infty\phi(z_i)\phi(\zeta_i)/\Phi(\zeta_i)dy_i$, for $i=1,\ldots,n$ such that $u_i=1$. 
Alternatively, a score residual based on all observations (including the censored ones) can be defined by 
\begin{eqnarray}\label{score_residual}
S^*_i=\dfrac{s_i-E(s_i)}{\sqrt{\mbox{Var}(s_i)}}=\dfrac{z_i-\dfrac{\rho_i}{\sqrt{1-\rho_i^2}}\dfrac{\phi(\zeta_{i})}{\Phi(\zeta_{i})}}{\sqrt{\Phi(\mu_{2i})\left(1+\mu_{2i}\rho_i^2\dfrac{\phi(\mu_{2i})}{\Phi(\mu_{2i})}+\dfrac{\rho_i^2}{1-\rho_i^2}\Psi_i\right)}},
\end{eqnarray}
for $i=1,\ldots,n$. In practice, we replace the unknown parameters by their maximum likelihood estimates. The evaluation of the goodness-of-fit of our proposed generalized Heckman model will be performed through the score residual analysis. Based on this approach, discrepant observations are identified, besides being possible to evaluate the existence of serious departures from the assumptions inherent to the model. If the model is appropriate, plots of residuals versus predicted values should have a random behavior around zero. Alternatively, a common approach is to build residual graphics with simulated envelopes \citep{atkinson1985}. In this case, it is not necessary to know about the distribution of the residuals, they just need to be within the region formed by the envelopes so indicating a good fit. Otherwise, residuals outside the envelopes are possible outliers or indicate that the model is not properly specified. We will apply the proposed score residual (\ref{score_residual}) to the MEPS data analysis. As will be shown, the residual analysis indicates that the normal assumption for the data is suitable in contrast with the non-robustness of the Heckman model mentioned in existing papers in the literature on sample selection models.

We now establish the consistency and asymptotic normality of the MLEs for our proposed generalized Heckman model. For this, we need to assume some usual regularity conditions.

\noindent (C1) The parameter space $\boldsymbol\Theta$ is closed and compact, and the true parameter value, say $\boldsymbol\theta_0$, is an interior point of $\boldsymbol\Theta$.

\noindent (C2) The covariates are a sequence of iid random vectors, and ${\bf F}_n$ is the information matrix conditional on the covariates. 

\noindent (C3) $E({\bf F}_n)$ is well-defined and positive definite, and  $E\left(\max_{\boldsymbol\theta\in\boldsymbol\Theta}\|{\bf F}_n\|\right)<\infty$, where $\|\cdot\|$ is the Frobenius norm. 

\begin{remark} Conditions (C2) and (C3) enable us to apply a multivariate Central Limit Theorem for iid random vectors for establishing the asymptotic normality of the MLEs. These conditions are discussed for instance in \cite{fahkau1985}.
\end{remark}

\begin{theorem}\label{asympt_results}
Under Conditions (C1)--(C3), the maximum likelihood estimator $\widehat{\boldsymbol\theta}$ of $\boldsymbol\theta$ for the generalized Heckman model is consistent and satisfies the weak convergence: ${\bf F}_n^{1/2}\left(\widehat{\boldsymbol\theta}-\boldsymbol\theta_0\right)\stackrel{d}{\longrightarrow}{\cal N}({\bf0},\bf{I})$, where $\bf{I}$ is the identity matrix, ${\bf F}_n$ is the conditional information matrix, and $\boldsymbol\theta_0$ denotes the true parameter vector value.
\end{theorem}
 
\begin{remark}
An important consequence of Theorem \ref{asympt_results} is that the classic Heckman model is regular under Conditions (C1)--(C3). Therefore, the MLEs for this model are consistent and asymptotically normal distributed, which is a new finding of this paper.
\end{remark}

\section{Monte Carlo Simulations}\label{simulations}

\subsection{Simulation design}
In this section, we develop Monte Carlo simulation studies to evaluate and compare the performance of the maximum likelihood estimators under the generalized Heckman, classic Heckman, Heckman-Skew, and Heckman-$t$ models when the assumption of either constant sample selection bias parameter or constant dispersion is not satisfied. To do this, six different scenarios with relevant characteristics for a more detailed evaluation were considered. In Scenarios 1 and 2, we use models with both varying dispersion and correlation (sample selection bias parameter) and with (I) exclusion restriction and (II) without the exclusion restriction. 

For Scenarios 3-6, the exclusion restriction is considered. More specifically, in Scenarios 3, 4, and 5, we have specified the following (III) constant dispersion and varying correlation; (IV) varying dispersion and constant correlation; (V) both constant dispersion and correlation. To evaluate the sensitivity in the parameter estimation of the selection models at high censoring, in Scenario 6 we simulated from the generalized Heckman model with (VI) varying both sample selection bias and dispersion parameters and an average censorship of $50\%.$

Scenario 1 aims to evaluate the performance of the generalized Heckman model and compare it with its competitors when the assumption of constant sample selection bias parameter and dispersion is not satisfied. Scenario 2 is devoted to demonstrating that despite the absence of exclusion restriction, our model can yield satisfactory parameter estimates. Scenarios 3 and 4 aim to justify the importance of modeling through covariates the correlation and dispersion parameters, respectively. Scenario 5 illustrates some problems that the generalized Heckman model can face as with the classic Heckman model. Finally, Scenario 6 was included to demonstrate the sensitivity of selection models to high correlation and high censoring. We here present the results from Scenario~1. The remaining results are presented in the Supplementary Material.

All scenarios were based on the following regression structures:
\begin{align}
\mu_{1i}&=\beta_{0}+\beta_{1}x_{1i}+\beta_{2}x_{2i},\label{eq_sel_heckman_psim1}  \\
\mu_{2i}&=\gamma_{0}+\gamma_{1}x_{1i}+\gamma_{2}x_{2i}+\gamma_{3}x_{3i},\label{eq_sel_heckman_psim2}\\
\log\sigma_{i}&=\lambda_{0}+\lambda_{1}x_{1i},\label{eq_sel_heckman_psim3}\\
\arctanh\,\rho_{i}&=\kappa_{0}+\kappa_{1}x_{1i},   \label{eq_sel_heckman_psim4}
\end{align}
for $i=1,\ldots,n$. All covariates were generated from a standard normal distribution and were kept constant throughout the experiment. The responses were generated from the generalized Heckman model according to each of the six configurations. We set the sample sizes $n=500,1000,2000$ and $N=1000$ Monte Carlo replicates. Pilot simulations showed that the choice of parameters used in the simulations does not affect the results, as long as they maintain the same average percentage of censorship.

We would like to highlight that there is no \texttt{R} package for fitting the Heckman-$t$ and Heckman skew-normal models. Therefore, we developed an \texttt{R} package (to be submitted) able to fit our proposed generalized Heckman model and also the sample selection models by \cite{marchenkoGenton} and \cite{ogundimu2016sample}.

In the maximization procedure to estimate parameters based on the $\texttt{optim}$ package, we consider as initial values for $\pmb{\beta}$, $\pmb{\gamma}$, $\lambda_{0}$, and $\kappa_{0}$ the maximum likelihood estimates by the classic Heckman model. For the remaining parameters, we set $\lambda_{i}=0$ and $\kappa_{j}=0$ for $i=1,\ldots,r$ and $j=1,\ldots,s$.  The initial values for the degrees of freedom of the Heckman-$t$ model and the skewness parameter for the Heckman-Skew model were set to $1$ and $2$, respectively. These values were chosen after some pilot simulations.

\subsection{Scenario 1: Varying sample selection bias and dispersion parameters}

 Here, we consider (\ref{eq_sel_heckman_psim1})-(\ref{eq_sel_heckman_psim4}) with $\pmb\beta=(1.1,0.7,0.1)^\top$, $\pmb\gamma=(0.9,0.5,1.1,0.6)^\top$, and also $\pmb\lambda=(-0.4,0.7)^\top$ and $\pmb\kappa=(0.3,0.5)^\top$.
 The regressors are kept fixed throughout the simulations with 
$x_{1i}, x_{2i}, x_{3i}\overset{iid}{\sim} \mathcal{N}(0,1)$ for all $i=1,\ldots,n.$ 

In Table \ref{2:res_cen3_tab1}, we present the empirical mean and root mean square error (RMSE) of the maximum likelihood estimates of the parameters based on the generalized Heckman, classic Heckman, Student-$t$, and skew-normal sample selection models under Scenario 1. From this table, we observe a good performance of the MLEs based on the generalized Heckman model, even for estimating the parameters related to the sample selection bias and dispersion. The bias and the RMSE under this model decrease for all the estimates as the sample size increases; therefore, suggesting the consistency of the MLEs, which is in line with our Theorem \ref{asympt_results}. On the other hand, even the regression structures for $\pmb\beta$ and $\pmb\gamma$ being correctly specified, we see that the MLEs for these parameters, based on the classic Heckman, skew-normal, and Student-$t$, do not provide satisfactory estimates even for a large sample. This illustrates the importance of considering covariates for the sample selection bias and dispersion parameters. The mean estimates of the degrees of freedom and skewness for the Student-$t$ and skew-normal sample selection models were $2.4$ and $0.8$, respectively.

\begin{sidewaysfigure}
\includegraphics[width=\textwidth]{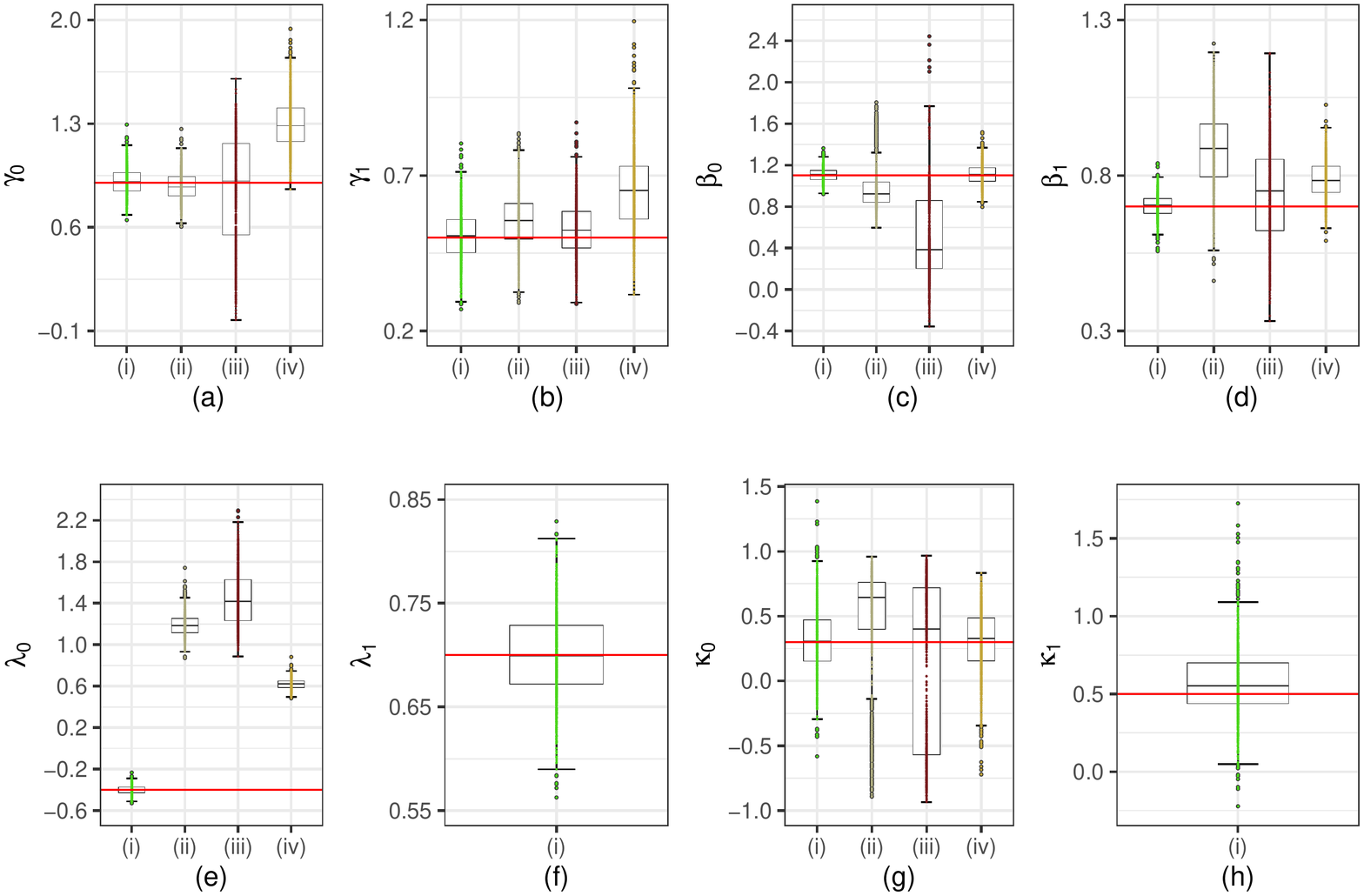}
\caption{Boxplots of the maximum likelihood estimates of the parameters (a) $\gamma_{0},$ (b) $\gamma_{1},$ (c) $\beta_{0},$ (d) $\beta_{1},$ (e) $\lambda_{0},$ (f) $\lambda_{1}$ and (g) $\kappa_{0}$ and (h) $\kappa_{1}$ based on the (i) generalized Heckman, (ii) classic Heckman, (iii) Heckman-Skew, and (iv) Heckman-$t$ sample selection models. Sample size $n=500.$} 
\label{2:cen3bp1}
\end{sidewaysfigure}

\begin{sidewaysfigure}
\includegraphics[width=\textwidth]{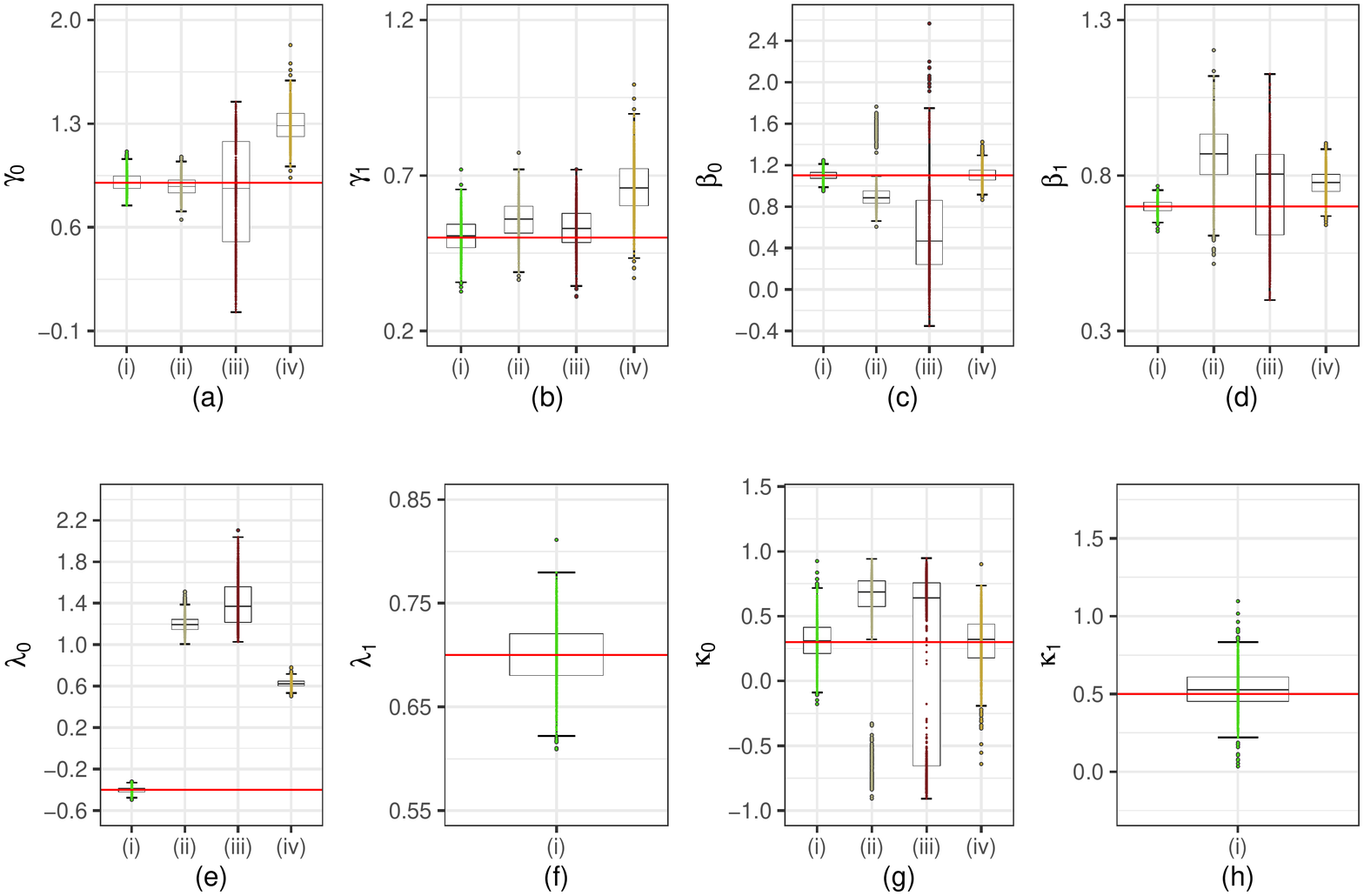}
\caption{Boxplots of the maximum likelihood estimates of the parameters (a) $\gamma_{0},$ (b) $\gamma_{1},$ (c) $\beta_{0},$ (d) $\beta_{1},$ (e) $\lambda_{0},$ (f) $\lambda_{1}$ and (g) $\kappa_{0}$ and (h) $\kappa_{1}$ based on the (i) generalized Heckman, (ii) classic Heckman, (iii) Heckman-Skew, and (iv) Heckman-$t$ sample selection models. Sample size $n=1000.$} 
\label{2:cen3bp2}
\end{sidewaysfigure}

\begin{sidewaysfigure}
\includegraphics[width=\textwidth]{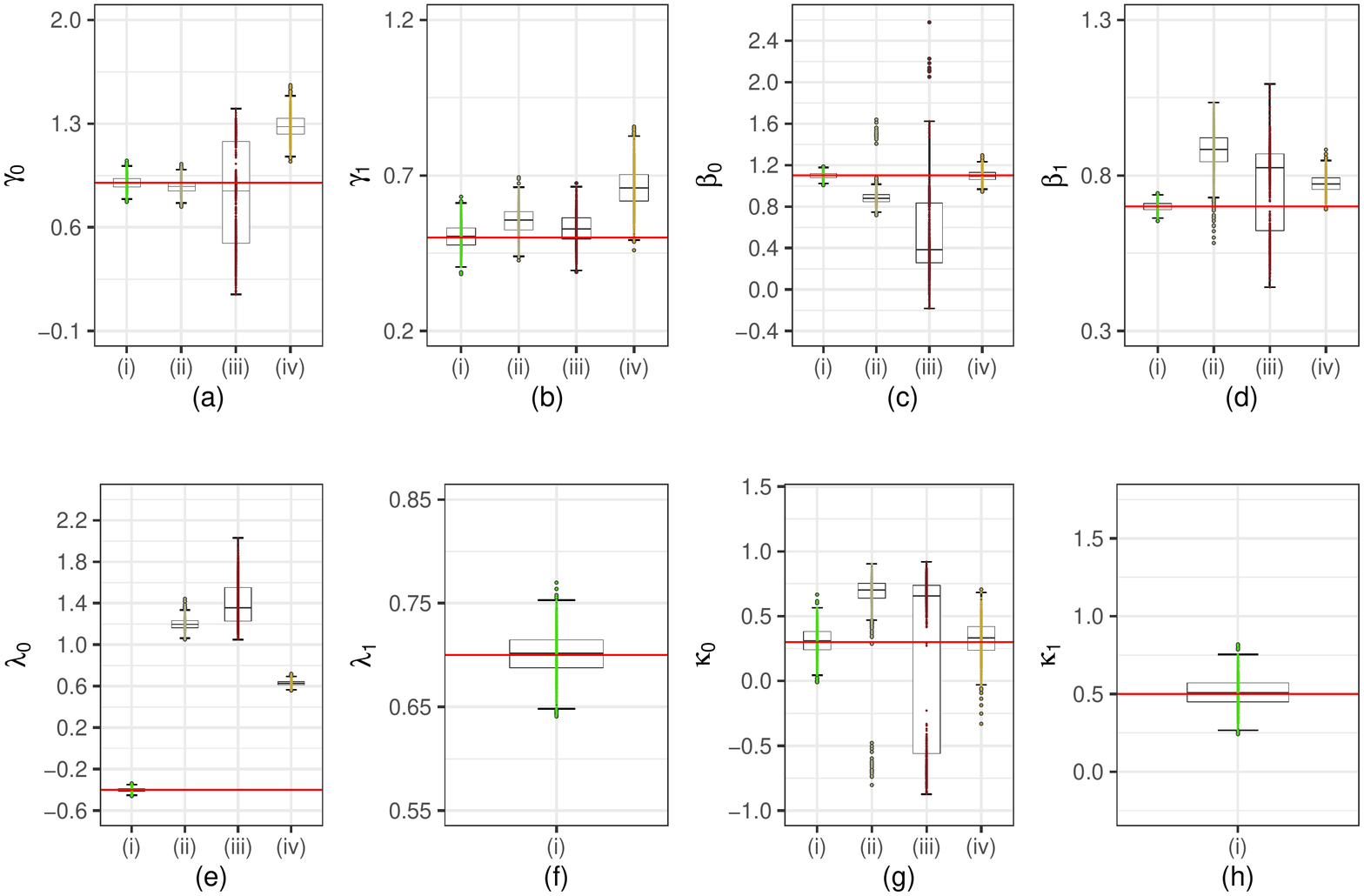}
\caption{Boxplots of the maximum likelihood estimates of the parameters (a) $\gamma_{0},$ (b) $\gamma_{1},$ (c) $\beta_{0},$ (d) $\beta_{1},$ (e) $\lambda_{0},$ (f) $\lambda_{1}$ and (g) $\kappa_{0}$ and (h) $\kappa_{1}$ based on the (i) generalized Heckman, (ii) classic Heckman, (iii) Heckman-Skew, and (iv) Heckman-$t$ sample selection models. Sample size $n=2000.$} 
\label{2:cen3bp3}
\end{sidewaysfigure}

The above comments are also supported by Figures \ref{2:cen3bp1}, \ref{2:cen3bp2}, and \ref{2:cen3bp3}, where boxplots of the parameter estimates are presented for sample sizes $n=500$, $n=1000$, and $n=2000$, respectively. We did not present the boxplots of the estimates of $\gamma_{1}$, $\gamma_{2}$, and $\beta_{1}$ since they behaved similarly to other boxplots.

\begin{table}
\centering
\caption{Empirical mean and root mean square error (RMSE) of the maximum likelihood estimates of the parameters based on the generalized Heckman, classic Heckman, Student-$t$, and skew-normal sample selection models under Scenario 1.} 
  \begin{tabular}{cccccc@{\hspace{-0.15ex}}ccc@{\hspace{-0.15ex}}ccc@{\hspace{-0.15ex}}cc}
  \hline \hline
  &&&\multicolumn{2}{c}{Generalized}&&\multicolumn{2}{c}{Classic}&&\multicolumn{2}{c}{Heckman}&&\multicolumn{2}{c}{Heckman} \\
  &&&\multicolumn{2}{c}{Heckman}&&\multicolumn{2}{c}{Heckman}&&\multicolumn{2}{c}{Skew-Normal}&&\multicolumn{2}{c}{Student-$t$} \\
  \cline{4-5} \cline{7-8} \cline{10-11}\cline{13-14} 
  \multicolumn{2}{c}{Parameters}&  $n$ & mean & RMSE&  & mean & RMSE &  & mean & RMSE & & mean & RMSE\\ \hline
  \hline
  \multirow{3}{*}{$\gamma_{0}$}& \multirow{3}{*}{0.9}&500      & 0.912 & 0.093 &  & 0.880 & 0.095 &  & 0.854 & 0.364 &  & 1.300 & 0.435   \\ 
  &&1000   & 0.905 & 0.063 &  & 0.878 & 0.069 &  & 0.820 & 0.364 &  & 1.293 & 0.409   \\ 
  &&2000   & 0.903 & 0.042 &  & 0.877 & 0.048 &  & 0.804 & 0.348 &  & 1.286 & 0.395   \\ 
  \hline                                                                                   
  \multirow{3}{*}{$\gamma_{1}$}& \multirow{3}{*}{0.5} &500     & 0.507 & 0.081 &  & 0.555 & 0.104 &  & 0.527 & 0.095 &  & 0.652 & 0.200   \\ 
  &&1000  & 0.506 & 0.056 &  & 0.558 & 0.085 &  & 0.530 & 0.075 &  & 0.663 & 0.185   \\ 
  &&2000  & 0.504 & 0.040 &  & 0.555 & 0.070 &  & 0.529 & 0.057 &  & 0.662 & 0.174   \\ 
  \hline                                                                                   
  \multirow{3}{*}{$\gamma_{2}$}& \multirow{3}{*}{1.1} &500     & 1.118 & 0.104 &  & 1.053 & 0.125 &  & 1.001 & 0.169 &  & 1.540 & 0.480   \\ 
  &&1000  & 1.109 & 0.074 &  & 1.046 & 0.099 &  & 0.978 & 0.162 &  & 1.522 & 0.441   \\ 
  &&2000  & 1.107 & 0.053 &  & 1.033 & 0.090 &  & 0.980 & 0.146 &  & 1.515 & 0.425   \\ 
  \hline                                                                                   
  \multirow{3}{*}{$\gamma_{3}$}& \multirow{3}{*}{0.6}&500      & 0.606 & 0.084 &  & 0.556 & 0.106 &  & 0.531 & 0.127 &  & 0.848 & 0.288   \\ 
  &&1000  & 0.604 & 0.065 &  & 0.539 & 0.098 &  & 0.504 & 0.130 &  & 0.828 & 0.253   \\ 
  &&2000   & 0.602 & 0.040 &  & 0.543 & 0.074 &  & 0.513 & 0.106 &  & 0.833 & 0.242   \\ 
  \hline                                                                                   
  \multirow{3}{*}{$\beta_{0}$}     & \multirow{3}{*}{1.1}  &500    & 1.105 & 0.068 &  & 0.998 & 0.261 &  & 0.496 & 0.734 &  & 1.112 & 0.103   \\ 
  &&1000 & 1.102 & 0.044 &  & 0.954 & 0.273 &  & 0.553 & 0.701 &  & 1.108 & 0.078   \\ 
  &&2000  & 1.099 & 0.029 &  & 0.892 & 0.230 &  & 0.498 & 0.697 &  & 1.096 & 0.051   \\ 
  \hline                                                                                    
  \multirow{3}{*}{$\beta_{1}$}     & \multirow{3}{*}{0.7}  &500    & 0.702 & 0.036 &  & 0.882 & 0.219 &  & 0.739 & 0.156 &  & 0.787 & 0.106   \\ 
  &&1000 & 0.701 & 0.020 &  & 0.860 & 0.191 &  & 0.753 & 0.158 &  & 0.777 & 0.087   \\ 
  &&2000  & 0.700 & 0.014 &  & 0.881 & 0.191 &  & 0.770 & 0.154 &  & 0.774 & 0.079   \\ 
  \hline                                                                                    
  \multirow{3}{*}{$\beta_{2}$}     & \multirow{3}{*}{0.1}  &500    & 0.098 & 0.048 &  & 0.140 & 0.196 &  & 0.040 & 0.225 &  & 0.101 & 0.078   \\ 
  &&1000 & 0.099 & 0.029 &  & 0.174 & 0.196 &  & 0.069 & 0.237 &  & 0.104 & 0.057   \\ 
  &&2000  & 0.100 & 0.019 &  & 0.221 & 0.147 &  & 0.105 & 0.193 &  & 0.113 & 0.042   \\ 
  \hline                                                                                   
  \multirow{3}{*}{$\lambda_{0}$}& \multirow{3}{*}{ $-$0.4} &500    & $-$0.400 & 0.042 &  & 0.174 & 0.577 &  & 0.366 & 0.771 &  & $-$0.476 & 0.113   \\ 
  &&1000   & $-$0.402 & 0.028 &  & 0.182 & 0.584 &  & 0.339 & 0.743 &  & $-$0.470 & 0.092   \\ 
  &&2000    & $-$0.401 & 0.019 &  & 0.182 & 0.582 &  & 0.331 & 0.734 &  & $-$0.463 & 0.075   \\ 
  \hline                                                                                   
  \multirow{3}{*}{$\lambda_{1}$}& \multirow{3}{*}{0.7}  &500      & 0.699 & 0.042 &  & $-$ & $-$ &  & $-$ & $-$ &  & $-$ & $-$    \\ 
  &&1000   & 0.700 & 0.031 &  & $-$ & $-$ &  & $-$ & $-$ &  & $-$ & $-$   \\ 
  &&2000    & 0.701 & 0.020 &  & $-$ & $-$ &  & $-$ & $-$ &  & $-$ & $-$    \\ 
  \hline                                                                                   
  \multirow{3}{*}{$\kappa_{0}$}  & \multirow{3}{*}{0.3 }  &500     & 0.314 & 0.248 &  & 0.507 & 0.688 &  & 0.133 & 0.846 &  & 0.307 & 0.300   \\ 
  &&1000   & 0.311 & 0.154 &  & 0.588 & 0.714 &  & 0.223 & 0.918 &  & 0.302 & 0.235   \\ 
  &&2000    & 0.309 & 0.106 &  & 0.811 & 0.608 &  & 0.307 & 0.814 &  & 0.336 & 0.158   \\ 
  \hline                                                                                   
  \multirow{3}{*}{$\kappa_{1}$}  & \multirow{3}{*}{0.5}   &500     & 0.573 & 0.240 &  & $-$ & $-$ &  & $-$ & $-$ &  & $-$ & $-$\\ 
  &&1000   & 0.531 & 0.135 &  & $-$ & $-$ &  & $-$ & $-$ &  & $-$ & $-$    \\ 
  &&2000    & 0.510 & 0.092 &  & $-$ & $-$ &  & $-$ & $-$ &  & $-$ & $-$    \\ 
  \hline
  \end{tabular}
\label{2:res_cen3_tab1}
\end{table}

We now provide some simulations to evaluate the size and power of likelihood ratio, gradient, and Wald tests. We consider Scenario 1 
and present the empirical significance level of the tests in Table \ref{2:tamanhoTesteCen1} for nominal significance levels at 1\%, 5\%, and 10\%.

Under the null hypothesis of absence of sample selection bias ($\rho_i=0$ for all $i$, that is $\kappa_0=\kappa_1=0$), the likelihood ratio, gradient, and Wald tests presented empirical values close to the nominal values only under the generalized Heckman model. For the other models, the type-I error was inflated and indicated the presence of selection bias. This suggests that the tests should be used with caution to test parameters of the sample selection models and that some confounding can be caused due to either varying sample selection bias or heteroskedasticity. It is important to point out that, even for the generalized Heckman model, the Wald test presents a considerable inflated type-I error for $n=500$.

\begin{table}
\centering
\caption{Empirical significance level of the likelihood ratio (LR), gradient (G) and Wald (W) tests for $H_{0}:\rho=0$.}
\begin{tabular}{c|ccccccccccccccccc}
\hline\hline
   &\multicolumn{3}{c}{Generalized}        &&\multicolumn{3}{c}{Classic}&& \multicolumn{3}{c}{Heckman}&& \multicolumn{3}{c}{Heckman} \\
       &\multicolumn{3}{c}{Heckman}          &&\multicolumn{3}{c}{Heckman} && \multicolumn{3}{c}{Skew-normal}       && \multicolumn{3}{c}{Student-$t$} \\
\cline{2-4} \cline{6-8} \cline{10-12} \cline{14-16}                        
$n$   &  LR  &  G  &  W  && LR & G  & W && LR & G  & W  && LR  & G & W \\ \hline \hline
                                        \cline{1-16}                                                                            
                           \multicolumn{16}{c}{$\alpha=1\%$}    \\ \hline \hline
$500   $  & 1.5&  1.1&  3.5&& 30.3&  1.6& 61.4&& 41.9&  6.4& 71.3&&  2.8&  1.5&  4.3\\
$1000$  & 0.9&  0.7&  2.6&& 72.4&  7.7& 92.7&& 82.7& 26.2& 95.6&&  3.4&  2.5&  4.0\\
$2000$  & 1.1&  0.7&  1.9&& 85.9& 16.5& 96.8&& 91.0& 52.4& 99.0&&  3.2&  2.6&  3.9\\
					\hline \hline
\multicolumn{16}{c}{$\alpha=5\%$}    \\ \hline \hline
$500   $  & 6.2&  5.6&  12.0 && 50.5& 12.5& 69.9&& 61.7& 26.9& 78.3&&  9.9&  7.7& 11.6\\
$1000$  & 4.9&  3.7&  6.9&& 86.9& 27.1& 95.5&& 89.9& 49.7& 97.8&& 10.1&  8.6& 11.8\\
$2000$  & 5.9&  5.1&  7.4&& 94.2& 40.8& 97.7&& 94.5& 68.3& 99.5&& 10.2&  9.4& 10.6\\
					\hline \hline
\multicolumn{16}{c}{$\alpha=10\%$}    \\ \hline \hline
$500   $  &  13.0& 10.5& 19.3&& 59.8& 25.2& 74.4&& 69.8& 40.8& 82.2&& 16.2& 14.8& 18.4\\
$1000$  & 9.4&  8.2& 13.7&& 91.5& 41.4& 96.1&& 93.3& 61.1& 98.2&& 16.9& 15.4& 18.0\\
$2000$  &11.2& 10.8& 12.6&& 96.3& 55.7& 98.3&& 95.5& 74.8& 99.6&& 16.1& 15.8& 16.4\\
					\hline \hline
\end{tabular}
\label{2:tamanhoTesteCen1}
\end{table}

In Table \ref{2:Poder1TesteCen1}, we present the empirical power of the likelihood ratio, gradient  and Wald tests (in percentage) for simulated data according to Scenario 1 under generalized Heckman, classic Heckman, Heckman-Skew, and Heckman-$t$ models, with significance level at $1\%, 5\%$ and $10\%$. From these results, we can observe that the tests, under the generalized Heckman model, provide high power, mainly when the sample size increases. On the other hand, since tests based on the other models do not provide the correct nominal significance level, the power of tests in these cases are not really comparable.

\begin{table}
\centering
\caption{Empirical power of the likelihood ratio (LR), gradient (G) and Wald (W) tests (in percentage) for simulated data according to Scenario 1 under generalized Heckman, classic Heckman, Heckman-Skew, and Heckman-$t$ models, with significance level at $1\%, 5\%$ and $10\%$.}
\begin{tabular}{c|ccccccccccccccccc}
\hline\hline
   &\multicolumn{3}{c}{Generalized}        &&\multicolumn{3}{c}{Classic}&& \multicolumn{3}{c}{Heckman}&& \multicolumn{3}{c}{Heckman} \\
       &\multicolumn{3}{c}{Heckman}          &&\multicolumn{3}{c}{Heckman} && \multicolumn{3}{c}{Skew-normal}       && \multicolumn{3}{c}{Student-$t$} \\
\cline{2-4} \cline{6-8} \cline{10-12} \cline{14-16}                        
$n$   &  LR  &  G  &  W  && LR & G  & W && LR & G  & W  && LR  & G & W \\ \hline \hline
                                        \cline{1-16}                                                                            
                           \multicolumn{16}{c}{$\alpha=1\%$}    \\ \hline \hline
$500   $  &71.7&68.4&75.5&&58.4& 8.9&81.9&&45.7& 4.5&79.7&&14.6& 9.3&18.4\\
$1000$  &95.3&95.2&96.9&&93.7&30.8&99.2&&  88.0&  14.0&98.4&&20.9&  16.0&26.6\\
$2000$  &99.9&99.9&99.9&&99.6&74.4& 100&&98.2&28.4&99.4&&48.3&  46.0&52.1\\
					\hline \hline
\multicolumn{16}{c}{$\alpha=5\%$}    \\ \hline \hline
$500   $  &88.8&87.8&91.6&&74.1&32.6&89.3&&69.2&20.1&86.3&&29.6&25.1&33.6\\
$1000$  &  99&98.9&99.4&&98.1&57.4&99.9&&94.5&33.5&  99&&41.9&37.9&46.2\\
$2000$  & 100& 100& 100&&99.9&  87.0& 100&&98.9&46.8&99.7&&68.8&66.9&70.9\\
					\hline \hline
\multicolumn{16}{c}{$\alpha=10\%$}    \\ \hline \hline
$500   $  &94.3&93.1&95.5&&82.4&48.4&91.9&&78.2&34.9&89.8&&39.8&37.1&43.1\\
$1000$  &99.8&99.8&99.8&&99.1&69.6& 100&&  96&45.5&  99&&52.4&49.4&54.7\\
$2000$  & 100& 100& 100&& 100&91.6& 100&&99.2&55.9&99.7&&78.4&77.8&79.4\\
					\hline \hline
\end{tabular}
\label{2:Poder1TesteCen1}
\end{table}

\section{MEPS Data Analysis}\label{3:application}

We present an application of the proposed model to a set of real data. Consider the outpatient expense data of the 2001 Medical Expenditure Panel Survey (MEPS) available in the \texttt{R} software in the package \texttt{ssmrob} \citep{ssmrob}. These data were also used by \cite{cameron2009microeconometrics}, \cite{marchenkoGenton}, and \cite{zhelonkin2013robustness} to fit the classic Heckman model, Heckman-$t$ model, and the robust version of the two-step method, respectively. The MEPS is a set of large-scale surveys of families, individuals, and their medical providers (doctors, hospitals, pharmacies, etc.) in the United States. It has data on the health services Americans use, how often they use them, the cost of these services, and how they are paid, as well as data on the cost and reach of health insurance available to American workers.

The sample is restricted to persons aged between 21 and 64 years and contains a variable response with $3,328$ observations of outpatient costs, of which 526 (15.8 \%) correspond to unobserved expenditure values identified as zero expenditure. It also includes the following explanatory variables: \texttt{Age} represents age measured in tens of years; \texttt{Fem} is an indicator variable for gender (female receives value 1); \texttt{Educ} represents years of schooling; \texttt{Blhisp} is an indicator for ethnicity (black or Hispanic receive a value of 1); \texttt{Totcr} is the total number of chronic diseases; \texttt{Ins} is the insurance status; and \texttt{Income} denotes the individual income.

\begin{figure}
\centering{
\includegraphics[width=18cm, height=6cm]{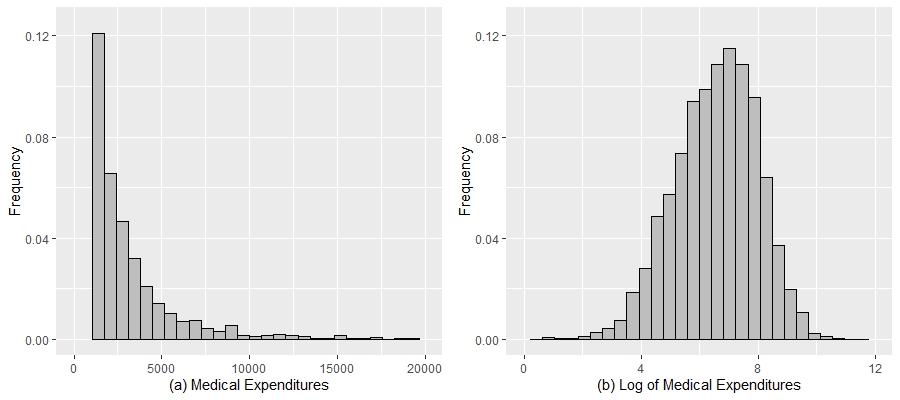}}
\caption{Histogram of medical expense data (a) and medical expense log (b).} 
\label{histogram}
\end{figure}

The variable of interest $Y_{1i}^{*}$ represents the log-expenditure on medical services of the $i$th individual. We consider the logarithm of the expenditure since it is highly skewed (i.e., see Figure~\ref{histogram}, where plots of the expenditure and log-expenditure are presented). The variable $Y_{2i}^{*}$ denoting the willingness of the $i$th individual to spend is not observed. We only observe $U_i=I\{Y_{2i}^{*}>0\}$, which represents the decision or not of the $i$th individual to spend on medical care.

According to \cite{cameron2009microeconometrics} and \cite{zhelonkin2013robustness}, it is natural to fit a sample selection model to such data, since the willingness to spend $(Y_{2}^{*})$ is likely to be related to the expense amount $(Y_{1}^{*}).$ However, after fitting the classic Heckman model and using the Wald statistics for testing $H_{0}:\rho=0$ against $H_{1}:\rho\neq 0,$ the conclusion is that there is no statistical evidence $(p\textrm{-value}>0.1)$ to reject $H_{0},$ that is, there is no sample selection bias. \cite{cameron2009microeconometrics} suspected this conclusion on the absence of sample selection bias, and \cite{marchenkoGenton} argued that a more robust model could evidence the presence of sample selection bias in the data; these authors proposed a Student-$t$ sample selection model to deal with this problem. However, as will be illustrated in this application, this problem of the classic Heckman model can be due to the assumption of constant sample selection bias and constant dispersion parameters rather than the normal assumption itself.

After a preliminary analysis, we consider the following regression structures for our proposed generalized Heckman model:
\begin{align*}
\mu_{1i}&= \beta_{0}+ \beta_{1}\texttt{Age}_{i}+ \beta_{2}\texttt{Fem}_{i}+ \beta_{3}\texttt{Educ}_{i}+ \beta_{4}\texttt{Blhisp}_{i}+\beta_{5}\texttt{Totchr}_{i}+\beta_{6}\texttt{Ins}_{i},  \\
\mu_{2i}&=\gamma_{0}+\gamma_{1}\texttt{Age}_{i}+\gamma_{2}\texttt{Fem}_{i}+\gamma_{3}\texttt{Educ}_{i}+\gamma_{4}\texttt{Blhisp}_{i}+\gamma_{5}\texttt{Totchr}_{i}+\gamma_{6}\texttt{Ins}_{i}+\gamma_{7}\texttt{Income}_{i}, \\
\log{\sigma_{i}} &=\lambda_{0}+\lambda_{1}\texttt{Age}_{i}+\lambda_{2}\texttt{Totchr}_{i}+\lambda_{3}\texttt{Ins}_{i},\\ 
\mbox{arctanh}\,\rho_{i}&=\kappa_{0}+\kappa_{1}\texttt{Fem}_{i}+\kappa_{2}\texttt{Totchr}_{i},
\end{align*}
for $i=1,\ldots, 3328$. The primary equation has the same covariates of the selection equation with the additional covariate \texttt{Income}, so that exclusion restriction is in force. In Table \ref{2:dados_reais_tab1}, we present the summary of the fits of the classic Heckman and generalized Heckman models. From this table, we can observe that the covariates \texttt{Fem} and \texttt{Totchr} are significant to explain the sample selection bias, by using any significance level. We perform a likelihood (LR) test for checking the absence ($H_0:\,\kappa_0=\kappa_1=\kappa_2=0$) or presence of sample selection bias. The LR statistic was $28.16$ with corresponding $p$-value equal to $3\times10^{-6}$. Therefore, our proposed generalized Heckman model is able to detect the presence of sample selection bias even under normal assumption. We also performed the gradient and Wald tests, which confirmed this conclusion. 

Further, the covariates \texttt{age}, \texttt{totcr}, and \texttt{ins} were significant for the dispersion parameter. For the selection equation, the covariate \texttt{Income} is not significant (significance level at 5\%) based on the generalized Heckman model in contrast with the classic Heckman model. Anyway, it is important to keep it in order to satisfy the exclusion restriction. Regarding the primary equation, we observe that the covariate \texttt{Ins} is only significant under the classic Heckman model. Another interesting point is that \texttt{Educ} is strongly significant for the primary equation under our proposed generalized Heckman model.

\begin{table}
\centering
\caption{Summary fits of the classic Heckman model (HM) and generalized Heckman model (GHM). The GHM summary fit contains estimates with their respective standard errors, z-value, $p\textrm{-value}$, and inferior and superior bounds of the $95\%$ confidence interval.} 
\begin{tabular}{lcccccccc}
\hline
	\multicolumn{9}{c}{\texttt{Selection Equation}}\\
	\hline
covariates & HM-est. & p-value & GHM-est. & stand. error & z-value & p-value & Inf. & Sup. \\
\hline
\texttt{Intercept} & $-$0.676 & 0.000 & $-$0.590 & 0.187 & $-$3.162 & 0.002 & $-$0.956 & $-$0.224 \\ 
\texttt{Age} & 0.088 & 0.001 & 0.086 & 0.026 & 3.260 & 0.001 & 0.034 & 0.138 \\ 
\texttt{Fem} & 0.663 & 0.000 & 0.630 & 0.060 & 10.544 & 0.000 & 0.513 & 0.747 \\ 
\texttt{Educ} & 0.062 & 0.000 & 0.057 & 0.011 & 4.984 & 0.000 & 0.035 & 0.079 \\ 
\texttt{Blhisp} & $-$0.364 & 0.000 & $-$0.337 & 0.060 & $-$5.644 & 0.000 & $-$0.454 & $-$0.220 \\ 
\texttt{Totchr} & 0.797 & 0.000 & 0.758 & 0.069 & 11.043 & 0.000 & 0.624 & 0.893 \\ 
\texttt{Ins} & 0.170 & 0.007 & 0.173 & 0.061 & 2.825 & 0.005 & 0.053 & 0.293 \\ 
\texttt{Income} & 0.003 & 0.040 & 0.002 & 0.001 & 1.837 & 0.066 & 0.000 & 0.005 \\ 
  \hline
	\multicolumn{9}{c}{\texttt{Primary Equation}}\\
	\hline
	covariates & HM-est. & p-value & GHM-est. & stand. error & z-value & p-value & Inf. & Sup. \\
  \hline 
\texttt{Intercept} & 5.044 & 0.000 & 5.704 & 0.193 & 29.553 & 0.000 & 5.326 & 6.082 \\ 
\texttt{Age} & 0.212 & 0.000 & 0.184 & 0.023 & 7.848 & 0.000 & 0.138 & 0.230 \\ 
\texttt{Fem} & 0.348 & 0.000 & 0.250 & 0.059 & 4.252 & 0.000 & 0.135 & 0.365 \\ 
\texttt{Educ} & 0.019 & 0.076 & 0.001 & 0.010 & 0.129 & 0.897 & $-$0.019 & 0.021 \\ 
\texttt{Blhisp} & $-$0.219 & 0.000 & $-$0.128 & 0.058 & $-$2.221 & 0.026 & $-$0.242 & $-$0.015 \\ 
\texttt{Totchr} & 0.540 & 0.000 & 0.431 & 0.031 & 14.113 & 0.000 & 0.371 & 0.490 \\ 
\texttt{Ins} & $-$0.030 & 0.557 & $-$0.103 & 0.051 & $-$1.999 & 0.046 & $-$0.203 & $-$0.002 \\ 
  \hline
	\multicolumn{9}{c}{\texttt{Dispersion Parameter}}\\
	\hline
	covariates & HM-est. & p-value & GHM-est. & stand. error & z-value & p-value & Inf. & Sup. \\
  \hline
\texttt{Intercept} & 1.271 & $-$ & 0.508 & 0.057 & 8.853 & 0.000 & 0.396 & 0.621 \\ 
\texttt{Age}    &$-$  &$-$  & $-$0.025 & 0.013 & $-$1.986 & 0.047 & $-$0.049 & 0.000 \\ 
\texttt{Totchr} &$-$  &$-$  & $-$0.105 & 0.019 & $-$5.475 & 0.000 & $-$0.142 & $-$0.067 \\ 
\texttt{Ins}    &$-$  &$-$  & $-$0.107 & 0.028 & $-$3.864 & 0.000 & $-$0.161 & $-$0.053 \\ 
  \hline
	\multicolumn{9}{c}{\texttt{Sample Selection Bias Parameter}}\\
	\hline
	covariates & HM-est. & p-value & GHM-est. & stand. error & z-value & p-value & Inf. & Sup. \\
  \hline
\texttt{Intercept} & $-$0.131 & 0.375 & $-$0.648 & 0.114 & $-$5.668 & 0.000 & $-$0.872 & $-$0.424 \\ 
\texttt{Fem}         & $-$ &$-$  & $-$0.403 & 0.136 & $-$2.973 & 0.003 & $-$0.669 & $-$0.137 \\ 
\texttt{Totchr}      & $-$ & $-$ & $-$0.438 & 0.186 & $-$2.353 & 0.019 & $-$0.803 & $-$0.073 \\ 
   \hline
\end{tabular}
\label{2:dados_reais_tab1}
\end{table}

\begin{figure}[H]
\includegraphics[width=18cm, height=6cm]{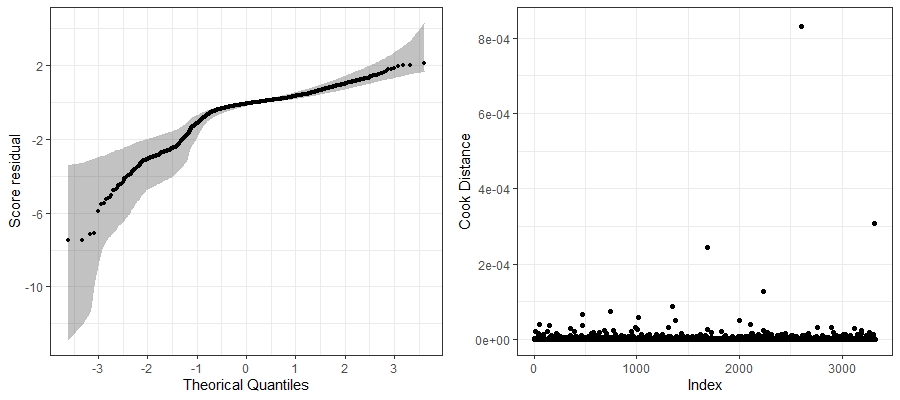}
\caption{QQ-plot and its simulated envelope for the score residuals (left) and index plot of the GCD (right) for the generalized Heckman model for the medical expenditure panel survey data.} 
\label{QQplot}
\end{figure}

We conclude this application by checking the goodness-of-fit of the fitted generalized sample selection Heckman model. In Figure \ref{QQplot}, we provide the QQ-plot of the score residuals given in (\ref{score_residual}) with simulated envelopes, and also a Cook distance plot for detecting global influence. Based on this last plot, we do not detect any outlier observation, since all points are below the reference line $2p/n=0.013$. Anyway, we investigate if the highlighted point \#2602 (above the line $8\times 10^{-4}$) is influential. We fitted our model by removing this observation and no changes either on the parameter estimates or different conclusions about the significance of covariates were obtained. Regarding the QQ-plot of the score residuals, we observe a great performance of our model, since 96\% of the points are inside of the envelope. This confirms that the normal assumption for this particular dataset is adequate and that our generalized Heckman model is suitable for the MEPS data analysis.


\section{Concluding Remarks}\label{concluding_remarks}

In this paper, a generalization of the Heckman model was proposed by allowing both sample selection bias and dispersion parameters to vary across covariates. We showed that the proposed model satisfies certain regularity conditions that ensure consistency and asymptotic normality of the maximum likelihood estimators. Furthermore, a proper score residual for sample selection models was proposed. These finding are new contributions on this topic. The MEPS data analysis based on the generalized Heckman model showed that the normal assumption for the data is suitable in contrast with existing findings in the literature. Future research should address (i) generalization of other sample selection models such as Student-$t$ and skew-normal to allow varying sample selection bias and dispersion parameters; (ii) proposal of proper residuals for other sample selection models; and (iii) deeper study of influence analysis. An \texttt{R} package for fitting our proposed generalized Heckman, Student-$t$, and skew-normal models has been developed and will be available soon. 

\section*{Appendix}

\noindent {\it Proof of Theorem \ref{regcond}}. We here show the results for the derivatives with respect to $\boldsymbol\beta$. The results involving the other derivatives follow similarly and therefore they are omitted.

For $i=1,\ldots,n$ and $j=1,\ldots,p$, we have that 
\begin{eqnarray}\label{scorebeta}
\dfrac{\partial\ell_i}{\partial\beta_j}=\left\{-\dfrac{\rho_i}{\sqrt{1-\rho_i^2}}\dfrac{\phi(\zeta_{i})}{\Phi(\zeta_{i})}+z_i\right\}x_{ij}u_i/\sigma_i.
\end{eqnarray}

By using basic properties of conditional expectation, it follows that $E\left(\dfrac{\partial\ell_i}{\partial\beta_j}\right)=E\left[E\left(\dfrac{\partial\ell_i}{\partial\beta_j}\big|U_i\right)\right]$ and it is immediate that $E\left(\dfrac{\partial\ell_i}{\partial\beta_j}\big|U_i=0\right)=0$. Let us now compute the conditional expectations involved in $E\left(\dfrac{\partial\ell_i}{\partial\beta_j}\big|U_i=1\right)$.

Here it is worth to remember the notations $z_i=\dfrac{y_i-\mu_{1i}}{\sigma_i}$ and $\zeta_i=\dfrac{\mu_{2i}+z_i\rho_i}{\sqrt{1-\rho_i^2}}$, for $i=1,\ldots,n$. We now use the conditional density function given in (\ref{1:dens_heckman}) to obtain that 
\begin{eqnarray*}
E\left(\dfrac{\phi(\zeta_{i})}{\Phi(\zeta_{i})}\big|U_i=1\right)&=&\dfrac{1}{\sigma_i\Phi(\mu_{2i})}\int_{-\infty}^\infty\phi(\zeta_i)\phi(z_i)dy_i=\dfrac{1}{2\pi\sigma_i\Phi(\mu_{2i})}\int_{-\infty}^\infty\exp\{-(\zeta_i^2+z_i^2)/2\}dy_i\\
&=&\dfrac{e^{-\mu_{2i}^2/2}}{2\pi\sigma_i\Phi(\mu_{2i})}\int_{-\infty}^\infty\exp\left\{-\dfrac{(y_i-\mu_{1i}+\sigma_i\mu_{2i}\rho_i)^2}{2\sigma_i^2(1-\rho_i^2)}\right\}dy_i=
\sqrt{1-\rho_i^2}\dfrac{\phi(\mu_{2i})}{\Phi(\mu_{2i})},
\end{eqnarray*}
where the last equality follows by identifying a normal kernel in the integral. On the other hand, we use the fact that $Z_i$ given $U_i=1$ has mean equal to $\mu_{1i}+\rho_i\sigma_i\dfrac{\phi(\mu_{2i})}{\Phi(\mu_{2i})}$ (more details are given in the Supplementary Material) and get
\begin{eqnarray*}
E\left(Z_i\big|U_i=1\right)=-\dfrac{\mu_{1i}}{\sigma_i}+\dfrac{1}{\sigma_i}E(Y_i|U_i=1)=-\dfrac{\mu_{1i}}{\sigma_i}+\dfrac{1}{\sigma_i}\left(\mu_{1i}+\rho_i\sigma_i\dfrac{\phi(\mu_{2i})}{\Phi(\mu_{2i})}\right)=\rho_i\dfrac{\phi(\mu_{2i})}{\Phi(\mu_{2i})}.
\end{eqnarray*}

With the results above, we obtain that $E\left(\dfrac{\phi(\zeta_{i})}{\Phi(\zeta_{i})}\big|U_i\right)=0$ almost surely and therefore $E\left(\dfrac{\partial\ell_i}{\partial\beta_j}\right)=0$ for $j=1,\ldots,p$. 

We now concentrate our attention to prove the identity stated in the theorem. It follows that
\begin{eqnarray*}
\dfrac{\partial^2\ell_i}{\partial\beta_j\beta_l}=-\left\{\dfrac{\rho_i^2}{1-\rho_i^2}\left[\zeta_i\dfrac{\phi(\zeta_i)}{\Phi(\zeta_i)}+\dfrac{\phi^2(\zeta_i)}{\Phi^2(\zeta_i)}\right]+1\right\}\dfrac{x_{ij}x_{il}}{\sigma_i^2}u_i
\end{eqnarray*}
and
\begin{eqnarray*}
\dfrac{\partial\ell_i}{\partial\beta_j}\dfrac{\partial\ell_i}{\partial\beta_l}=\left\{z_i^2-2z_i\dfrac{\rho_i}{\sqrt{1-\rho_i^2}}\dfrac{\phi(\zeta_i)}{\Phi(\zeta_i)}+\dfrac{\rho_i^2}{1-\rho_i^2}\dfrac{\phi^2(\zeta_i)}{\Phi^2(\zeta_i)}\right\}\dfrac{x_{ij}x_{il}}{\sigma_i^2}u_i,
\end{eqnarray*}
where we have used that $u_i^2=u_i$ (since $u_i\in\{0,1\}$) in the last equality. It is immediate that $E\left(\dfrac{\partial^2\ell_i}{\partial\beta_j\beta_l}\big|U_i=0\right)=-E\left(\dfrac{\partial\ell_i}{\partial\beta_j}\dfrac{\partial\ell_i}{\partial\beta_l}\big|U_i=0\right)=0$.

Following in a similar way as before, after some algebra we obtain that
\begin{eqnarray*}
E\left(\zeta_i\dfrac{\phi(\zeta_i)}{\Phi(\zeta_i)}\big|U_i=1\right)=-\mu_{2i}(1-\rho_i^2)\dfrac{\phi(\mu_{2i})}{\Phi(\mu_{2i})}\quad\mbox{and}\quad
E\left(Z_i^2\big|U_i=1\right)=1-\mu_{2i}\rho_i^2\dfrac{\phi(\mu_{2i})}{\Phi(\mu_{2i})}.
\end{eqnarray*}

By combining these results, we have that 
\begin{eqnarray*}
-E\left(\dfrac{\partial^2\ell_i}{\partial\beta_j\beta_l}\big|U_i=1\right)&=&E\left(\dfrac{\partial\ell_i}{\partial\beta_j}\dfrac{\partial\ell_i}{\partial\beta_l}\big|U_i=1\right)\\
&=&\left\{1+\mu_{2i}\rho_i^2\dfrac{\phi(\mu_{2i})}{\Phi(\mu_{2i})}+\dfrac{\rho_i^2}{1-\rho_i^2}E\left(\dfrac{\phi^2(\zeta_i)}{\Phi^2(\zeta_i)}\big|U_i=1\right)\right\}\dfrac{x_{ij}x_{il}}{\sigma_i^2}.
\end{eqnarray*}

Since the conditional expectations coincide, the marginal expectations also coincide so giving the desired result. $\square$

\noindent {\it Proof of Theorem \ref{asympt_results}}. Conditions (C1)--(C3) and Theorem \ref{regcond} give us the consistency of the MLEs. To establish the asymptotic normality of the estimators, we need to show that the third derivatives of the log-likelihood function are bounded by integrable functions not depending on the parameters. 

We will show here that this is possible for the derivatives involving the $\boldsymbol\beta$'s. The other cases follow in a similar way as discussed in the proof of Theorem \ref{regcond} and therefore they are omitted. 

By computing the third derivatives with respect to the $\boldsymbol\beta$'s and using the triangular inequality, we have that 
\begin{eqnarray*}
\left|\dfrac{\partial^3\ell_i}{\partial\beta_j\partial\beta_l\partial\beta_k}\right|&\leq&\dfrac{\rho_i^2}{\sigma_i^3(1-\rho_i^2)^{3/2}}\left\{(1+z_i^2)\dfrac{\phi(\zeta_i)}{\Phi(\zeta_i)}+\zeta_i^2\dfrac{\phi^2(\zeta_i)}{\Phi^2(\zeta_i)}+2\zeta_i\dfrac{\phi^2(\zeta_i)}{\Phi^2(\zeta_i)}+2\dfrac{\phi^2(\zeta_i)}{\Phi^3(\zeta_i)}\right\}x_{ij}x_{il}x_{ik}\\
&\equiv& g_i(\boldsymbol{\theta})\leq g_i(\boldsymbol{\theta}^*),
\end{eqnarray*}
for $j,l,k=1,\ldots,p$, where $\boldsymbol{\theta}^*=\mbox{argmax}_{\boldsymbol{\theta}\in\boldsymbol\Theta}g_i(\boldsymbol{\theta})$, which is well-defined due to Assumption (C1).

We now need to show that the expectations of the terms in $g_i(\boldsymbol{\theta}^*)$ are finite. Let us show that $E_{\boldsymbol\theta_0}\left(\zeta_i^{*2}\dfrac{\phi^2(\zeta_i^*)}{\Phi^2(\zeta_i^*)}\right)<\infty$, where $E_{\boldsymbol\theta_0}(\cdot)$ denotes the expectation with respect to the true parameter vector value $\boldsymbol\theta_0$ and $\zeta_i^*$ is defined as $\zeta_i$ by replacing $\boldsymbol{\theta}$ by $\boldsymbol{\theta}^*$. The proofs for the remaining terms follow from this one or in a similar way. 

For $\zeta_i^*\leq\sqrt2$, it follows that $\phi^2(\zeta_i^*)/\Phi^2(\zeta_i^*)\leq \left\{2\pi\Phi^2(\sqrt2)\right\}^{-1}$. Now, consider $\zeta_i^*>\sqrt2$. Theorem 1.2.6 from \cite{durrett} gives us the following inequality for $x>0$:
\begin{eqnarray*}
\left(\dfrac{1}{x}-\dfrac{1}{x^3}\right)e^{-x^2/2}\leq \int_x^\infty e^{-y^2/2}dy.
\end{eqnarray*}

Using this inequality and under $\zeta_i^*>\sqrt2$, we obtain that 
$\dfrac{\phi^2(\zeta_i^*)}{\Phi^2(\zeta_i^*)}\leq \dfrac{\zeta_i^{*3}}{\zeta_i^{*2}-1}\leq \zeta_i^{*}$. These results imply that
\begin{eqnarray*}
E_{\boldsymbol\theta_0}\left(\zeta_i^{*2}\dfrac{\phi^2(\zeta_i^*)}{\Phi^2(\zeta_i^*)}\right)&=&E_{\boldsymbol\theta_0}\left(\zeta_i^{*2}\dfrac{\phi^2(\zeta_i^*)}{\Phi^2(\zeta_i^*)}I\left\{\zeta_i^*\leq\sqrt2\right\}\right)+E_{\boldsymbol\theta_0}\left(\zeta_i^{*2}\dfrac{\phi^2(\zeta_i^*)}{\Phi^2(\zeta_i^*)}I\left\{\zeta_i^*>\sqrt2\right\}\right)\\
&\leq& \sqrt{2}\left\{2\pi\Phi^2(\sqrt2)\right\}^{-1}+E_{\boldsymbol\theta_0}\left(\left|\zeta_i^{*}\right|^3\right)<\infty,
\end{eqnarray*}
with $E_{\boldsymbol\theta_0}\left(\left|\zeta_i^{*}\right|^3\right)<\infty$ being proved in the same lines that the first two moments presented in the Supplementary Material, which completes the proof of the desired result. $\square$


\begin{thebibliography}{100}



\bibitem[{Arabmazar and Schmidt(1981)}]{arabmazar1981further} 
\textsc{Arabmazar, A. and Schmidt, P.} (1981).  
\newblock{Further evidence on the robustness of the tobit estimator to heteroskedasticity}.
\newblock{Journal of Econometrics} \textbf{17}, 253--258.

\bibitem[{Atkinson(1985)}]{atkinson1985} 
\textsc{Atkinson, A. C.} (1985).  
\newblock{Plots, Transformations, and Regression}.
\newblock{Oxford: Oxford University Press}.


\bibitem[{Azzalini et al.(2019)}]{azzalinietal} 
\textsc{Azzalini, A. and Kim, H. M. and Kim, H. J.} (2019).  
\newblock{Sample selection models for discrete and other non-Gaussian response variables}.
\newblock{Statistical Methods and Applications} \textbf{19}, 27--56.

\bibitem[{Cameron and Trivedi(2009)}]{cameron2009microeconometrics} 
\textsc{Cameron, C. A. and Trivedi, P. K.} (2009).  
\newblock{Microeconometrics Using Stata}.
\newblock{TX: Stata Press}.




\bibitem[{Chib et al.(2009)}]{Siddhartha} 
\textsc{Chib, S. and Greenberg, E. and Jeliazkov, I.} (2009).  
\newblock{Estimation of semiparametric models in the presence of endogeneity and sample selection}.
\newblock{Journal of Computational and Graphical Statistics} \textbf{18}, 321--348.


\bibitem[{Cook(1977)}]{RDCook} 
\textsc{Cook, R. D.} (1977).  
\newblock{Detection of influential observation in linear regression}.
\newblock{Technometrics} \textbf{19}, 15--18.


\bibitem[{Cook and Weisberg(1982)}]{cook1982residuals} 
\textsc{Cook, R.D. and Weisberg, S.} (1982).  
\newblock{Residuals and Influence in Regression}.
\newblock{New York: Chapman and Hall}.



\bibitem[{Cox and Hinkley(1979)}]{cox1979theoretical} 
\textsc{Cox, D.R. and Hinkley, D.V.} (1979).  
\newblock{Theoretical Statistics}.
\newblock{Chapman and Hall/CRC}.



\bibitem[{Donald(1995)}]{donald1995two} 
\textsc{Donald, S.G.} (1995).  
\newblock{Two-Step estimation of heteroskedastic sample selection models}.
\newblock{Journal of Econometrics} \textbf{65}, 347--380.



\bibitem[{Durrett(2019)}]{durrett} 
\textsc{Durrett, R.} (2019).  
\newblock{Probability: Theory and Examples}.
\newblock{Cambridge University Press}.


\bibitem[{Enders(2010)}]{enders2010applied} 
\textsc{Enders, C.K.} (2010).  
\newblock{Applied Missing Data Analysis}.
\newblock{Guilford Press: New York, NY}.



\bibitem[{Fahrmeir and Kaufmann(1985)}]{fahkau1985} 
\textsc{Fahrmeir, L. and Kaufmann, H.} (1985).  
\newblock{Consistency and asymptotic normality of the maximum likelihood estimator in generalized linear models}.
\newblock{Annals of Statistics} \textbf{13}, 342--368.


\bibitem[{Heckman(1974)}]{heckman1974} 
\textsc{Heckman, J.J.} (1974).  
\newblock{Shadow prices, market wages, and labor supply}.
\newblock{Econometrica} \textbf{42}, 679--694.



\bibitem[{Heckman(1976)}]{heckman1976} 
\textsc{Heckman, J.J.} (1976).  
\newblock{The common structure of statistical models of truncation, sample selection and limited dependent variables and a simple estimator for such models}.
\newblock{Annals of Economic and Social Measurement} \textbf{5}, 475--492.


\bibitem[{Heckman(1979)}]{heckman1979} 
\textsc{Heckman, J.J.} (1979).  
\newblock{Sample selection bias as a specification error}.
\newblock{Econometrica} \textbf{47}, 153--161.


\bibitem[{Hurd(1979)}]{hurd1979estimation} 
\textsc{Hurd, M.} (1979).  
\newblock{Estimation in truncated samples when there is heteroscedasticity}.
\newblock{Journal of Econometrics} \textbf{11}, 247-258.


\bibitem[{Kim et al.(2019)}]{Taeyoung} 
\textsc{Kim, H., Roh, T. and Choi, T.} (2019).  
\newblock{Bayesian analysis of semiparametric Bernstein polynomial regression models for data with sample selection}.
\newblock{Statistics} \textbf{53}, 1--30.



\bibitem[{Lai and Tsay(2018)}]{TsayPin} 
\textsc{Lai, H.P. and Tsay, W.J.} (2018).  
\newblock{Maximum simulated likelihood estimation of the panel sample selection model}.
\newblock{Econometric Reviews} \textbf{37}, 744--759.


\bibitem[{Leung and Yu(1996)}]{leung1996choice} 
\textsc{Leung, S.F. and Yu, S.} (1996).  
\newblock{On the choice between sample selection and two-part models}.
\newblock{Journal of Econometrics} \textbf{72}, 197--229.



\bibitem[{Leung and Yu(2000)}]{Leung2000} 
\textsc{Leung, S. F. and Yu, S.} (2000).  
\newblock{Collinearity and two-step estimation of sample selection models: Problems, origins, and remedies}.
\newblock{Computational Economics} \textbf{15}, 173--199.



\bibitem[{Marchenko and Genton(2012)}]{marchenkoGenton} 
\textsc{Marchenko, Y. V. and Genton, M. G.} (2012).  
\newblock{A Heckman selection-t model}.
\newblock{Journal of the American Statistical Association} \textbf{107}, 304--317.



\bibitem[{Marra and Wyszynski(2016)}]{MARRA2016110} 
\textsc{Marra, G. and Wyszynski, K.} (2016).  
\newblock{Semi-parametric copula sample selection models for count responses}.
\newblock{Computational Statistics and Data Analysis} \textbf{104}, 110-129.


\bibitem[{Mu and Zhang(2018)}]{Beili} 
\textsc{Mu, B. and Zhang, Z.} (2018).  
\newblock{Identification and estimation of heteroscedastic binary choice models with endogenous dummy regressors}.
\newblock{Econometrics Journal} \textbf{21}, 218--246.



\bibitem[{Newey(2009)}]{Newey2009} 
\textsc{Newey, W.K.} (2009).  
\newblock{Two-step series estimation of sample selection models}.
\newblock{Econometrics Journal} \textbf{12}, 217--229.


\bibitem[{Ogundimu and Hutton(2016)}]{ogundimu2016sample} 
\textsc{Ogundimu, E.O. and Hutton, J.L.} (2016).  
\newblock{A sample selection model with skew-normal distribution}.
\newblock{Scandinavian Journal of Statistics} \textbf{43}, 172--190.



\bibitem[{Powell(1986)}]{powell1986symmetrically} 
\textsc{Powell, J.L.} (1986).  
\newblock{Symmetrically trimmed least squares estimation for tobit models}.
\newblock{Econometrica} \textbf{54}, 1435--1460.



\bibitem[{R Core Team(2020)}]{SoftwareR} 
\textsc{R Core Team} (2020).  
\newblock{R: A Language and Environment for Statistical Computing}.
\newblock{R Foundation for Statistical Computing}.



\bibitem[{Wiesenfarth and Kneib(2010)}]{Wiesenfarth} 
\textsc{Wiesenfarth, M. and Kneib, T.} (2010).  
\newblock{Bayesian geoadditive sample selection models}.
\newblock{Journal of the Royal Statistical Society - Series C} \textbf{59}, 381--404.




\bibitem[{Wojtys et al.(2018)}]{Giampiero} 
\textsc{Wojtys, M. and Marra, G. and Radice, R.} (2018).  
\newblock{Copula based generalized additive models for location, scale and shape with non-random sample selection}.
\newblock{Computational Statistics and Data Analysis} \textbf{127}, 1--14.


\bibitem[{Wyszynski and Marra(2018)}]{Wysmar2018} 
\textsc{Wyszynski, K. and Marra, G.} (2018).  
\newblock{Sample selection models for count data in R}.
\newblock{Computational Statistics} \textbf{33}, 1385-1412.



\bibitem[{Xie and Wei(2007)}]{XIE20074692} 
\textsc{Xie, F.-C. and Wei, B.-W.} (2007).  
\newblock{Diagnostics analysis for log-Birnbaum–Saunders regression models}.
\newblock{Computational Statistics and Data Analysis} \textbf{51}, 4692--4706.

\bibitem[{Zhelonkin et al.(2014)}]{ssmrob} 
\textsc{Zhelonkin, M. and Genton, M. G. and Ronchetti, E.} (2014).  
\newblock{R package ssmrob: Robust estimation and inference in sample selection models}.
\newblock{https://CRAN.R-project.org/package=ssmrob}.


\bibitem[{Zhelonkin et al.(2016)}]{zhelonkin2013robustness} 
\textsc{Zhelonkin, M. and Genton, M. G. and Ronchetti, E.} (2016).  
\newblock{Robust inference in sample selection models}.
\newblock{Journal of the Royal Statistical Society - Series B} \textbf{78}, 805--827.

\end{thebibliography}

\end{document}